\documentclass{moriond}

\usepackage{amssymb,amsmath}
\newcommand{\be}{\begin{eqnarray}}

\newcommand{\ee}{\end{eqnarray}}

\newcommand{\BTT}{\ensuremath{ \langle BTT \rangle}}
\newcommand{\BB}{\ensuremath{ \langle BB \rangle}}
\newcommand{\BTE}{\ensuremath{ \langle BTE \rangle}}
\newcommand{\BEE}{\ensuremath{ \langle BEE \rangle}}
\newcommand{\Bell}{\ensuremath{\boldsymbol\ell}}

\newcommand{\bs}{\boldsymbol}

\usepackage{hyperref}

\newcommand{\Photo}{}

\begin{document}
\vspace*{4cm}
\title{The Holiest Grail}

\author{ P.~Daniel~Meerburg}

\address{CITA, University of Toronto, 60 St. George Street, Toronto, Canada}

\maketitle\abstracts{
I will discuss to what degree the cosmic microwave background (CMB) can be used to constrain primordial non-Gaussianity involving one tensor and two scalar fluctuations, focusing on the correlation of one $B$-mode polarization fluctuation with two temperature fluctuations (BTT). In the simplest models of inflation, the tensor-scalar-scalar primordial bispectrum is non-vanishing and is of the same order in slow-roll parameters as the scalar-scalar-scalar bispectrum. I will show that constraints from an experiment like CMB-Stage IV using this observable are more than an order of magnitude better than those on the same primordial coupling obtained from temperature measurements alone. I will argue that $B$-mode non-Gaussianity opens up an as-yet-unexplored window into the early Universe, demonstrating that significant information on primordial physics remains to be harvested from CMB anisotropies. BTT presents a measure of both primordial tensors and primordial non-Gaussianity, two of the most sought after signatures of the inflationary paradigm. 
}

\section{Introduction}

Our primordial Universe can be described by only 2 parameters; the amplitude $A_s$ and tilt $n_s$ of the power of initial fluctuations. Although beautiful in its simplicity, from a empirical point of view, it limits the amount of information we can hope to extract from the early Universe, which by itself is our only hope of ever probing energy scales way beyond the standard model of particle physics. Currently, our best best bet on further exploring the early Universe in search for other degrees of freedom is Cosmic Microwave Background polarization and Large Scale Structure. In this short note I will discuss a potentially very interesting observable that is sensitive to non-Gaussianities generated by the interaction between a single graviton and two scalars. The bispectrum associated with this interaction, leaves an imprint on the on the CMB in the temperature bispectrum \cite{Shiraishi2010}. However, similar to the tensor constraint derived from the $TT$ power spectrum, the tensor contribution to the measured spectrum of fluctuations, has large cosmic variance from the scalar contribution. Therefore, we propose to correlate one $B$ mode with tow $T$ modes, i.e. $\BTT$. The replacement of one leg with a $B$ mode, significantly lowers cosmic variance and allows us to constrain this type of interaction with increasing precision as $B$ mode measurements improve. 

\section{Graviton-Scalar-Scalar Bispectrum}
As a starting point we consider a primordial Universe with a single field driving an inflationary expansion (SFSR). For such model, the bispectrum of a graviton interacting with two scalars was first computed in \cite{Maldacena2002} and is given by
\be
\left\langle \zeta(\vec{k}_1)\zeta(\vec{k}_2)h^{\pm}(\vec{k}_3) \right\rangle = (2\pi)^3 F^{\pm200}(\vec{k}_1,\vec{k}_2,\vec{k}_3) \delta \left(\sum_{n=1}^3\vec{k}_n\right), 
\ee
where we have defined
\be
F^{\pm200}(\vec{k}_1,\vec{k}_2,\vec{k}_3) &=&  \frac{H_*^4}{4M_{{\rm pl}}^4\epsilon_*} I(k_1,k_2,k_3)e_{ab}^{\mp}(\bs{k}_3)\bs{k}_1^a \bs{k}_2^b, 
\label{eq:FTensScal}
\ee
with
\be
\mathcal{I}(k_1,k_2,k_3) &=& \frac{1}{k_1^3k_2^3k_3^3}\left(-k_t + \frac{k_1k_2+k_2k_3+k_1k_3}{k_t} + \frac{k_1k_2k_3}{k_t^2}\right) \, .
\ee
The shape is specific for SFSR and can change if alternative models are considered. As a test-case we consider the shape above. It is very similar to a local shape. In a more general form, we  define
\be
\langle \zeta(\bs{k}_1)\zeta(\bs{k}_2)h^{\pm}(\bs{k}_3) \rangle &=& (2\pi)^3 16 \pi^4 A_s^2 \sqrt{r}f_\mathrm{NL}^{h\zeta\zeta}  \delta^{(3)} \left(\sum_{n=1}^3\bs{k}_n\right) \mathcal{I}(k_1,k_2,k_3) e_{ab}^{\mp}(\bs{k}_3)\bs{k}_1^a \bs{k}_2^b, 
\ee
For SFSR one can show that $f^{h \zeta \zeta}_{\rm NL} = \sqrt{r}/16$. $\mathcal{I}$ can be replaced by any shape that has is symmetric in $k_1$ and $k_2$. For a scale invariant spectrum of perturbations $\mathcal{I} \propto k^{-8}$. 


Because tensors mostly affect transverse modes, a graviton-scalar-scalar bispectrum will get most of its contribution in the squeezed configuration (local-type). In projection, the signal in the CMB is maximized for large angle $B$-modes ($\ell_B$) and small $T$ modes ($\ell_T^1,\ell_T^2$). Because of the symmetry properties of the $BTT$ correlation function, in the limit the $T$-modes are equal in length (i.e. when $\ell_T^1 = \ell_T^2$) the correlation function vanishes \cite{Meerburg2016}.

The flat-sky bispectrum can be computed through
\be
B_{\Bell_1 \Bell_2 \Bell_3}^{TTB} &=& 16 \pi^2 A_s^2 \sqrt{r} f_\mathrm{NL}^{h\zeta\zeta} (\Bell_1 \times \Bell_3)\times \int dk_1^z dk_2^z~\mathcal{I}\left(k_1^R,k_2^R,k_3^R\right) \nonumber\\
&& \Delta_{T}^{\zeta} (k_1^z, \ell_1)\Delta_{T}^{\zeta} (k_2^z, \ell_1)\Delta_{P}^h (k_3^z, \ell_3) \frac{\sqrt{2}k_3^z}{k_3^R \ell_3^2}\left[ k_1^z (\Bell_2 \cdot \Bell_3) - k_2^z (\Bell_1 \cdot \Bell_3)\right], \label{eq:BTTbispectrum2}
\ee
where the integrals over $k_1^z, k_2^z$ run from $-\infty$ to $+\infty$, $k_3^z \equiv -(k_1^z + k_2^z)$, and $k_i^R \equiv \sqrt{(k_i^z)^2 + (\ell_i/D_R)^2}$, with $D_R = \tau_0- \tau_R$, where $\tau_R$ is the conformal time at the peak of the CMB visibility function, and, for $X = T, P$ and $s = \zeta, h$,
\be
\Delta_{X}^s (k_i^z,\ell_i) \equiv \int_0^{\tau_0} \frac{d\tau }{D^2} S_X^s(k_i^R)e^{-i k^i_z \tilde{D}}, \label{eq:Delta_Xs}
\ee
with $\tilde{D}= \tau -\tau_R$.

We have computed the $\langle BTT\rangle$ correlation function in the flat-sky limit $\ell_{\rm min} = 10$ in a Planck fiducial cosmology. We would again like to stress that the shape considered is a test-case; its predicted amplitude is too low to be observed (similar to the scalar bispectrum from SFSR). 

\section{Signal to Noise}
In the flat-sky approximation the signal-to-noise ratio is given by the following integral over multipoles (rather than a discrete sum in the full-sky case) \cite{HU2000SN}: 
\be
\left(\frac{S}{N}\right)^2 = \frac{f_{\rm sky}}{4\pi^3} \int d^2\Bell_2 \int d^2 \Bell_3  \frac{\left(B_{\left(-\Bell_2 - \Bell_3\right),  \Bell_2, \Bell_3}^{TTB}\right)^2}{\mathcal{C}^{TT}_{\ell_1} \mathcal{C}^{TT}_{\ell_2} \mathcal{C}^{BB}_{\ell_3} }, 
\label{eq:2DSN}
\ee
where $\mathcal{C}_{\ell} = C_{\ell} + N_{\ell}$ with $C_{\ell}$ is the angular power spectrum, $N_{\ell}$ the noise, and we integrated out the $\Bell_1$ direction. 

\begin{figure}[t] 
   \centering
   \includegraphics[width=2.5in]{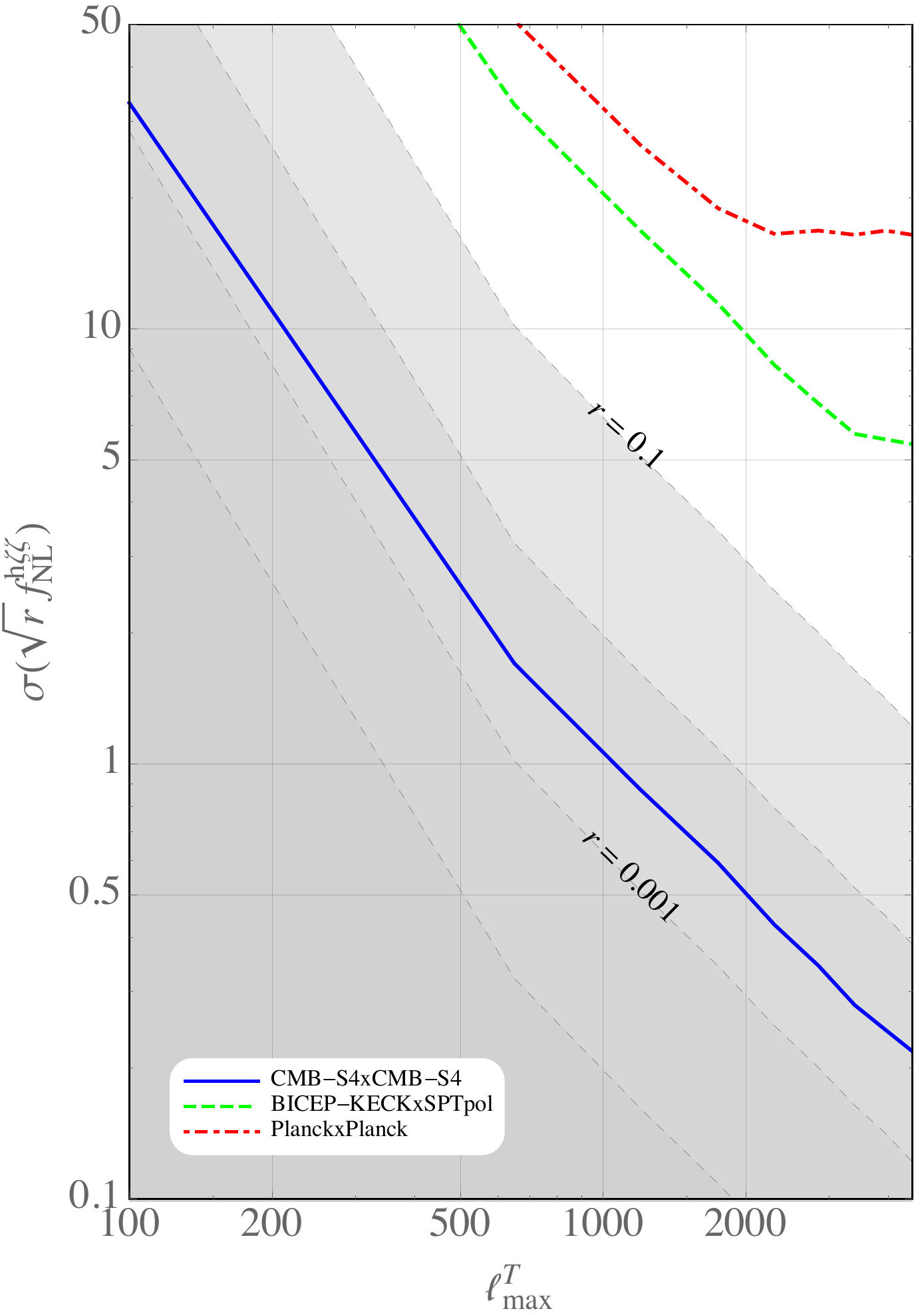} 
   \includegraphics[width=2.5in]{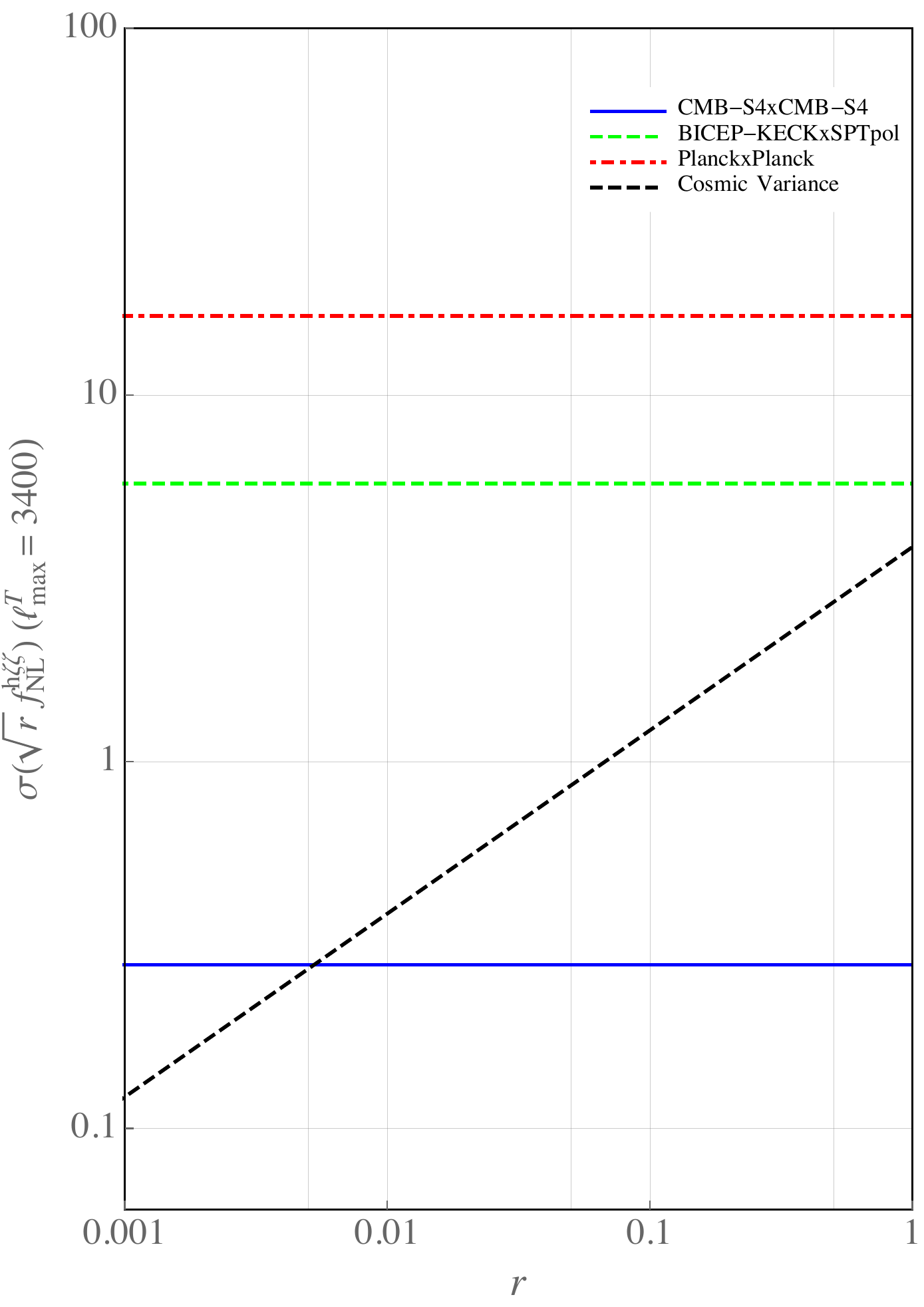} 
   \caption{Left: Forecasts for $\sigma (\sqrt{r} f_{\rm NL}^{h\zeta\zeta})$ for various CMB experiments as a function of $\ell_{\rm max}$. The colored lines present constraints when cosmic variance is negligible. The figure shows that cosmic variance would be subdominant for current and near future experiments if $r=0.01$. For an experiment like CMB-Stage IV the total variance would be dominated by cosmic variance and not by instrumental noise unless $r \lesssim 0.001$ (with $\ell^T_{\rm max} = 4500$). Right: Forecasts for $\sigma(\sqrt{r} f_{\rm NL}^{h\zeta\zeta})$ for various CMB experiments. This figure illustrates that current experiments are all noise dominated for allowed values of $r$. CMB-Stage IV is comic variance dominated unless $r \lesssim 0.005$ (with $\ell^T_{\rm max} = 3400$).  Cosmic variance limit can only be reduced if we consider more modes, i.e. by increasing $\ell_{\rm max}^T$).}
   \label{fig:deltafnl}
\end{figure}

The forecasted $1\sigma$ constraints on $\sqrt{r}f^{h \zeta \zeta}_{\rm NL}$ are plotted in Fig.~\ref{fig:deltafnl} for several experiments including CMB-Stage IV. We assumed a noise dominated $B$-mode. If $r$ is detected, the grey band indicates how the constraints are shifted. The plot should be compared to constraints from $TTT$ on the same interaction, which $\sqrt{r} f^{h \zeta \zeta}_{\rm NL} \sim \mathcal{O}(10)$ \cite{Maresuke2011}. CMB-Stage IV can improve those constraints by almost 2 orders of magnitude. Note that $\BTE$ and $\BEE$ should lead to similar constraints. Other combinations are expected to be worse (e.g. $\langle TEE \rangle$) since they are currently already cosmic variance dominated on large scales. Improvements can be achieved, but only from small angle $T$ and $E$ modes. 

\section{Discussion and Conclusion}

We have explored the potential of the $\langle BTT \rangle$ bispectrum as probe of the early universe. The odd intrinsic parity of $B$-modes gives this bispectrum some properties which differ from that of the $\langle TTT \rangle$ bispectrum, but both are generically non-vanishing in a parity-conserving universe, and are sourced by primordial bispectra which are predicted to be of the same order in slow-roll parameters in single-field slow-roll inflation.

One advantage of the $\langle BTT \rangle$ bispectrum is that the signal suffers less from cosmic variance than its $\langle TTT \rangle$ counterpart for constraining the tensor-scalar-scalar bispectrum.  Our analysis shows that with this observable it should be possible to constrain the level of non-Gaussianity to $\sigma(\sqrt{r}f_{\rm NL}^{h\zeta\zeta}) \sim \mathcal{O}(0.1) f_{\rm sky}^{-1/2}$. 
  
We considered $\langle BTT \rangle$ bispectrum motivated by single-field slow-roll inflation, which maximizes when large-scale $B$ modes are correlated with small-scale $T$ modes.  This shape has some experimental advantages, since a search for such a bispectrum could be performed, for instance, by cross-correlating a map of $B$-modes on large scales from a current or upcoming ground-based CMB experiment with small-scale $T$ fluctuations, such as those measured with the Planck satellite.  On the other hand, it would be useful to consider other shapes for the $\langle BTT \rangle$ bispectrum, perhaps motivated by specific early-universe models.

We have not included possible contamination from dust, systematics, or lensing.  Dust is a well-known contaminant in estimates of the \BB\ power spectrum on large scales at the frequencies probed with ground-based experiments.   Estimates of the \BTT\ bispectrum will in principle be sensitive to correlations between large-scale dust polarization and small-scale dust intensity; while this may be less of an issue than in the power spectrum measure, this needs to be investigated in future work.  Similarly, instrumental systematics which affect the measure of $B$ on large scales should be decoupled from those that affect $T$ on small scales, making this analysis less sensitive to systematics than the  \BB\ power spectrum.  Finally, lensing converts $E$-mode polarization to $B$-mode polarization.  As with measurements of the \BB\ power spectrum, delensing to reduce effective noise from lensing needs to be performed when measuring the \BTT\ bispectrum.  In a universe with primordial gravitational waves, the \BTT\ bispectrum will also contain a non-primordial signal on very large scales arising from correlations between $B$-modes from Thomson scattering after reionization and the curl mode of CMB lensing, which affects pairs of temperature modes. This is analogous to the  scattered $E$-mode--lensing correlation induced by scalars in estimates of the $\langle ETT \rangle$ bispectrum   \cite{lewis2011}.

We have focused on the $\langle BTT \rangle$ correlation function. However, other combinations sensitive to the coupling between scalars and tensors will add to the total signal-to-noise ratio. In particular, $\langle BEE \rangle$ and $\langle BTE \rangle$ are expected to have similar constraining power. 
In addition, similar to the use of $E$-modes for the scalar bispectrum, the $B$ and $E$ modes are projected through functions that have different nulls, which improves the mapping from the primordial space.
In summary, the $\langle BTT \rangle$ bispectrum and other non-Gaussian correlations involving $B$-modes open up a new window into the early Universe.  Ongoing and future CMB experiments will naturally make observations which allow us to carry out searches for and place non-trivial constraints on primordial tensor non-Gaussianity.  While more theoretical work remains to discover the full value of $B$-mode non-Gaussianity, this new set of observables has the potential to be a very rich set of tools for probing primordial physics.

\section*{Acknowledgments}
I would like to thank Joel Meyers, Alex van Engelen and Yacine Ali-Haimoud for their contribution to this work.

\section*{References}

\bibliography{proceedings}

\end{document}